\documentclass[12pt]{article}
\usepackage{graphicx}
\usepackage{amsfonts}

\newcommand{\RR}{{\mathbb R}}

\newlength{\extraspace}
\newlength{\extraspaces}
\newcommand{\be}{\begin{equation}
\addtolength{\abovedisplayskip}{\extraspaces}
\addtolength{\belowdisplayskip}{\extraspaces}
\addtolength{\abovedisplayshortskip}{\extraspace}
\addtolength{\belowdisplayshortskip}{\extraspace}}
\newcommand{\ee}{\end{equation}}

\newcommand{\ba}{\begin{eqnarray}
\addtolength{\abovedisplayskip}{\extraspaces}
\addtolength{\belowdisplayskip}{\extraspaces}
\addtolength{\abovedisplayshortskip}{\extraspace}
\addtolength{\belowdisplayshortskip}{\extraspace}}
\newcommand{\ea}{\end{eqnarray}}

\newcommand{\newsection}[1]{
\vspace{12mm}
\pagebreak[3]
\addtocounter{section}{1}
\setcounter{equation}{0}
\setcounter{subsection}{0}
\setcounter{footnote}{0}
\noindent{\bf \thesection. #1}
\nopagebreak
\medskip
\nopagebreak}

\def\Box{{\hbox{$\sqcup$}\llap{\hbox{$\sqcap$}}}}

\begin{document}
\addtolength{\baselineskip}{1.5mm}

\thispagestyle{empty}
\begin{flushright}
hep-th/0307188\\
\end{flushright}
\vbox{}
\vspace{2.5cm}

\begin{center}
{\LARGE{Black diholes in five dimensions
        }}\\[16mm]
{Edward Teo}
\\[6mm]
{\it Department of Physics,
National University of Singapore, 
Singapore 119260}\\[15mm]

\end{center}
\vspace{2cm}

\centerline{\bf Abstract}\bigskip
\noindent
Using a generalized Weyl formalism, we show how stationary, axisymmetric 
solutions of the four-dimensional vacuum Einstein equation can be turned 
into static, axisymmetric solutions of five-dimensional dilaton gravity
coupled to a two-form gauge field. This procedure is then used to obtain 
new solutions of the latter theory describing pairs of extremal magnetic 
black holes with opposite charges, known as black diholes. These diholes 
are kept in static equilibrium by membrane-like conical singularities 
stretching along two different directions. We also present solutions 
describing diholes suspended in a background magnetic field, and with 
unbalanced charges.


\newpage

\newsection{Introduction}

The Bonnor dipole solution in Einstein--Maxwell theory has been known
for some time \cite{Bonnor}, although it was not until recently that
Emparan \cite{Emparan} found it could be interpreted as a pair
of extremal Reissner--Nordstr\"om black holes with magnetic charges of 
equal magnitude but opposite signs. To maintain the black holes in static 
equilibrium, there are in general conical singularities, interpreted as 
struts or cosmic strings, pushing or pulling on the two black holes. 
Alternatively, they can be held apart by introducing a background
magnetic field. Emparan called such a configuration a `black dihole', 
and in his paper examined a number of their properties.

The analog of the Bonnor dipole in Kaluza--Klein theory was found
by Gross and Perry \cite{Gross}. They obtained it by Euclideanizing
the Kerr solution $t\rightarrow ix^5$, adding on a flat time direction,
and then dimensionally reducing along $x^5$. It was subsequently 
generalized to a solution of Einstein--Maxwell--dilaton theory with a 
general dilaton coupling in \cite{Davidson,Galtsov}. (A way to derive this
solution using the Weyl formalism can be found in \cite{Liang}.) These 
dilatonic dihole solutions can be interpreted as pairs of extremal 
dilatonic black holes with opposite charges.

It is possible to generalize the dihole solution in various ways. For 
example, a dihole carrying unbalanced charges in Einstein--Maxwell--dilaton 
theory was derived in \cite{Liang}. Such a solution carries a net charge, 
unlike those considered previously. In \cite{Emparan_Teo}, the non-extremal 
generalization of the dihole solution was constructed, and its various 
properties, such as its thermodynamics and the interaction between the 
black holes, were studied. Furthermore, the solution was embedded into 
string / M-theory, and a microscopic description of the entropy of a certain
near-extremal dihole was found in terms of an effective string model 
consisting of interacting strings and anti-strings.

It would be of interest, particularly from the viewpoint of string or 
M-theory, to construct diholes in dimensions $D>4$, and there has 
been a few attempts in this direction. One possibility would be to start 
from the higher-dimensional Euclidean Kerr solution \cite{Myers} and add 
a flat time direction to it. However, this does not give a pair of 
Kaluza--Klein black holes upon dimensional reduction as one might 
hope, but rather a single spherical $(D-4)$-brane 
\cite{Dowker:1996,Janssen,Emparan:2001}. In $D=5$, for example, one 
obtains a Kaluza--Klein string wound in a circular loop. Clearly, a 
different approach is needed to find higher-dimensional dihole solutions. 

In a separate development, the Weyl formalism was recently generalized
by Emparan and Reall \cite{Emparan_Reall} to arbitrary dimensions
$D\geq4$. Space-times belonging to the generalized Weyl class admit
$D-2$ orthogonal commuting Killing vectors, and are specified by $D-3$
axisymmetric solutions to the Laplace equation in flat three-dimensional 
space. A particular member of the generalized Weyl class is the 
five-dimensional Schwarzschild black hole. Moreover, it was pointed out 
in \cite{Emparan_Reall} that solutions describing superpositions of 
Schwarzschild black holes can also be obtained within this formalism. 
A five-dimensional generalization of the Israel--Khan solution, describing 
a collinear array of Schwarzschild black holes, was subsequently constructed 
in \cite{Tan}. A charged version of this solution was also found,
with each black hole having a fixed mass-to-charge ratio.

The fact that it is possible to construct solutions describing multiple
black holes in five dimensions (possibly carrying charges of the same sign), 
strongly suggests that it would also be possible to find a 
five-dimensional analog of the dihole solution. Indeed, the aim of 
this paper is to present dihole solutions of five-dimensional 
dilaton gravity coupled to a two-form gauge field, following the
strategy of \cite{Liang}. As will be described in Sec.~2, the derivation 
makes use of the generalized Weyl formalism of \cite{Emparan_Reall}. 
A formal similarity is observed between the {\it four\/}-dimensional 
Ernst equation and the field equations coming from five-dimensional 
dilaton gravity theory when a generalized Weyl symmetry is assumed. This 
will allow us to turn known solutions of one system into new solutions of 
the other. Indeed, starting from the four-dimensional Kerr solution, we 
will show in Sec.~3 how this gives the desired five-dimensional dihole 
solution. We will also show how to rewrite this solution in a simpler 
form using C-metric-type coordinates.

In Sec.~4, we will discuss this solution in some detail.
We show how the individual extremal dilaton black holes may be recovered,
and that there are in general conical singularities in the space-time.
These conical singularities are actually two-dimensional membranes 
extending along two different directions, similar to the situation 
in the two-black hole configuration of \cite{Tan}. Instead of using conical 
singularities to achieve equilibrium, it is possible to use a background 
magnetic field to do so. We show how this is done by applying a 
five-dimensional analog of the Harrison transformation to the solution.

In Sec.~5, we derive a five-dimensional dihole solution carrying
unbalanced charges following the strategy of \cite{Liang}. This involves
applying the solution-generating technique to the Kerr solution with a NUT 
parameter \cite{Demianski}, instead of the usual Kerr solution. 
We then briefly analyze some of its properties. The paper ends off 
with a discussion of some possible avenues for future research.

\newsection{Solution-generating technique}

We shall consider five-dimensional gravity coupled to a dilaton field
$\phi$ and two-form gauge field $B_{ab}$, with the action
\be
\label{action}
\int{\rm d}^5x\sqrt{-g}\left(R-\frac{1}{2}\partial_a\phi\partial^a\phi
-\frac{1}{12}{\rm e}^{\alpha\phi}H_{abc}H^{abc}\right).
\ee
Here, $H_{abc}\equiv\partial_{a}B_{bc}+\partial_{b}B_{ca}+\partial_{c}B_{ab}$ 
is the three-form field strength, and $\alpha$ is a parameter 
governing the coupling of the dilaton to the gauge field. 
Since changing the sign of $\alpha$ is equivalent to changing the
sign of $\phi$, it is sufficient to consider only $\alpha\geq0$.
The case $\alpha=2\sqrt{\frac{2}{3}}$ is particularly important as it 
arises from the low-energy effective action of string theory
when compactified to five dimensions.
By varying $g_{ab}$, $\phi$ and $B_{ab}$, we obtain from (\ref{action})
the respective field equations:
\ba
\label{Rab}
&&R_{ab}=\frac{1}{2}\partial_a\phi\partial_b\phi
+\frac{1}{4}{\rm e}^{\alpha\phi}\left(H_{acd}H_b{}^{cd}
-\frac{2}{9}g_{ab}H_{cde}H^{cde}\right),\\
&&\Box\phi-\frac{\alpha}{12}{\rm e}^{\alpha\phi}H_{abc}H^{abc}=0\,,\\
&&\nabla_a({\rm e}^{\alpha\phi}H^{abc})=0\,.
\ea

We seek a solution to these equations that is static and axisymmetric
with respect to two angular coordinates $\varphi$ and $\psi$, i.e., an
$\RR\times U(1)\times U(1)$ symmetry. The most general line element 
satisfying these conditions can be written as
\be
\label{metric}
{\rm d}s^2=-f\,{\rm d}t^2+l\,{\rm d}\varphi^2+k\,{\rm d}\psi^2+{\rm e}^{\mu}({\rm d}\rho^2+{\rm d}z^2)\,,
\ee
where $f$, $k$, $l$ and $\mu$ are functions of $\rho$ and $z$ only.
This is an immediate generalization of the Weyl form
for static, axisymmetric space-times in four dimensions \cite{Emparan_Reall}. 
Furthermore, we assume that the only non-zero component of the 
two-form gauge field is $B_{\varphi\psi}\equiv B$. Both $B$ and 
$\phi$ are also functions of $\rho$ and $z$ only.

Now, we begin with three of the six non-trivial equations coming 
from (\ref{Rab}):
\ba
\label{Rtt}
2{\rm e}^\mu DR_{tt}&=&(Df_\rho)_\rho+(Df_z)_z
-Df^{-1}(f_\rho^2+f_z^2)
=\frac{2}{3}\frac{Df}{kl}{\rm e}^{\alpha\phi}(B_\rho^2+B_z^2)\,,\\
-2{\rm e}^\mu DR_{\varphi\varphi}&=&(Dl_\rho)_\rho+(Dl_z)_z
-Dl^{-1}(l_\rho^2+l_z^2)
=-\frac{1}{3}\frac{D}{k}{\rm e}^{\alpha\phi}(B_\rho^2+B_z^2)\,,\\
\label{Rpsipsi}
-2{\rm e}^\mu DR_{\psi\psi}&=&(Dk_\rho)_\rho+(Dk_z)_z
-Dk^{-1}(k_\rho^2+k_z^2)
=-\frac{1}{3}\frac{D}{l}{\rm e}^{\alpha\phi}(B_\rho^2+B_z^2)\,,
\ea
where we have defined $D^2\equiv fkl$, and denoted 
$f_\rho\equiv\partial f/\partial\rho$, $f_z\equiv\partial f/\partial z$, etc.,
for brevity.
Considering the combination ${\rm e}^\mu D^{-1}(klR_{tt}-fkR_{\varphi\varphi}
-flR_{\psi\psi})$, we obtain
\be
D_{\rho\rho}+D_{zz}=0\,.
\ee
A solution is $D=\rho$. Substituting this back into 
(\ref{Rtt}) and (\ref{Rpsipsi}) respectively, we obtain
\ba
\label{feqn}
&f_{\rho\rho}+f_{zz}+\rho^{-1}f_\rho
-f^{-1}(f_\rho^2+f_z^2)={2\over3}\rho^{-2}f^2
{\rm e}^{\alpha\phi}(B_\rho^2+B_z^2)\,,&\\
\label{etaeqn}
&k_{\rho\rho}+k_{zz}+\rho^{-1}k_\rho
-k^{-1}(k_\rho^2+k_z^2)=-{1\over3}\rho^{-2}fk
{\rm e}^{\alpha\phi}(B_\rho^2+B_z^2)\,.&
\ea
Also, the dilaton and two-form gauge field equations respectively become
\ba
\label{phieqn}
&\phi_{\rho\rho}+\phi_{zz}+\rho^{-1}\phi_\rho={1\over2}\alpha\rho^{-2}f{\rm e}^{\alpha\phi}(B_{\rho}^2+B_z^2)\,,&\\
\label{Beqn}
&B_{\rho\rho}+B_{zz}-\rho^{-1}B_\rho=-f^{-1}(B_\rho f_\rho+B_zf_z)-\alpha(\phi_\rho B_\rho+\phi_zB_z)\,.&
\ea

If we define 
\be
\label{ft}
\tilde f^2\equiv f{\rm e}^{\alpha\phi},\qquad
w\equiv i\sqrt{4+3\alpha^2\over12}\,B\,, 
\ee
then (\ref{feqn}), (\ref{phieqn}) and (\ref{Beqn}) imply that
\ba
\label{fteqn}
&\tilde f(\tilde f_{\rho\rho}+\tilde f_{zz}+\rho^{-1}\tilde f_\rho)
-\tilde f_\rho^2-{\tilde f_z}^2+\rho^{-2}\tilde f^4(w_\rho^2+w_z^2)=0\,,&\\
\label{weqn}
&\tilde f(w_{\rho\rho}+w_{zz}-\rho^{-1}w_\rho)+2(w_\rho\tilde f_\rho+w_z\tilde f_z)=0\,.&
\ea
These two equations are equivalent to the Ernst equation (c.f.~Eqn.~(2.12a) 
and (2.12b) of \cite{Islam}) coming from the four-dimensional vacuum 
Einstein equation, if we consider a stationary, axisymmetric 
line element of the form
\be
\label{5DWeyl}
{\rm d}s^2=-\tilde f({\rm d}t-w\,{\rm d}\varphi)^2+\rho^2\tilde f^{-1}{\rm d}\varphi^2+{\rm e}^{\tilde \mu}({\rm d}\rho^2+{\rm d}z^2)\,.
\ee
This is the crucial correspondence
which would allow us to obtain solutions to the action (\ref{action}),
starting from solutions to pure Einstein gravity. Note, however, that this 
correspondence in general gives a $B$ that is imaginary. To obtain a
real expression for $B$, the original solution must admit a parameter 
which can be analytically continued so that $w$ becomes imaginary
while leaving $\tilde f$ real. 

Supposing we have found suitable solutions $\tilde f$ and $w$ to the equations
(\ref{fteqn}) and (\ref{weqn}), the next step is to solve for $k$ and
$\phi$. (The latter would then give us $f$ by the first equation of
(\ref{ft}).) Note, from (\ref{phieqn}) and (\ref{fteqn}), that
\be
\label{phi}
\phi=\frac{6\alpha}{4+3\alpha^2}\,\ln\tilde f\,,
\ee
up to the addition of a harmonic function $\tilde{\phi}$, satisfying 
$\tilde{\phi}_{\rho\rho} + \tilde{\phi}_{zz} + \rho^{-1}\tilde{\phi}_\rho =0$.
For the choice $\tilde{\phi}=0$, we have
\be
\label{f}
f=\tilde f^\frac{8}{4+3\alpha^2}.
\ee
Similarly, we deduce from (\ref{etaeqn}) and (\ref{phieqn}) that
\be
k={\rm e}^{h-\frac{2\phi}{3\alpha}}={\rm e}^h\tilde f^{-\frac{4}{4+3\alpha^2}},
\ee
where $h$ is another arbitrary harmonic function. A suitable choice for
$h$ will be made below.

Having obtained $\phi$, $k$, $f$ and $B$, the final step is to solve 
for $\mu$. Using the three remaining equations of (\ref{Rab}),
namely those for $R_{\rho\rho}$, $R_{\rho z}$ and $R_{zz}$, we obtain
\ba
\label{mu_rho}
\rho^{-1}\mu_\rho&=&\hbox{$1\over2$}f^{-2}(f_\rho^2-f_z^2)
+\hbox{$1\over2$}k^{-2}(k_\rho^2-k_z^2)
+\hbox{$1\over2$}f^{-1}k^{-1}(f_\rho k_\rho-f_zk_z)
-\rho^{-1}f^{-1}f_\rho-\rho^{-1}k^{-1}k_\rho\cr
&&+\hbox{$1\over2$}(\phi_\rho^2-\phi_z^2)
+\hbox{$1\over2$}\rho^{-2}f{\rm e}^{\alpha\phi}(B_\rho^2-B_z^2)\,,\\
\label{mu_z}
\rho^{-1}\mu_z&=&f^{-2}f_\rho f_z+k^{-2}k_\rho k_z
+\hbox{$1\over2$}f^{-1}k^{-1}(f_\rho k_z+f_zk_\rho)
-\rho^{-1}f^{-1}f_z-\rho^{-1}k^{-1}k_z\cr
&&+\phi_\rho\phi_z+\rho^{-2}f{\rm e}^{\alpha\phi}B_\rho B_z\,.
\ea
These two equations can be integrated to obtain $\mu$. This therefore 
completes the procedure whereby a stationary, axisymmetric solution 
(\ref{5DWeyl}) to the four-dimensional vacuum Einstein equation may be 
converted into a static, axisymmetric solution of five-dimensional 
dilaton gravity coupled to a two-form gauge field. It is the analog 
of the result found in \cite{Liang} in four dimensions.

\newsection{Derivation of dihole solution}

A natural starting point would be the four-dimensional Kerr solution
with mass $m$ and angular momentum $a$, after performing the analytic 
continuation $a\rightarrow ia$. It turns out that this solution is most 
conveniently written in terms of prolate spheroidal coordinates $(p,q)$ 
as~\cite{Islam,Kramer}
\be
\label{4DKerr}
\tilde f=\frac{\sigma^2p^2-a^2q^2-m^2}{(\sigma p+m)^2-a^2q^2}\,,\qquad 
w=\frac{2ima(\sigma p+m)(1-q^2)}{\sigma^2p^2-a^2q^2-m^2}\,,
\ee
where $\sigma\equiv\sqrt{m^2+a^2}$. If necessary, it can be rewritten 
in terms of Weyl coordinates $(\rho,z)$ using the relations
\be
\label{prolate}
p=\frac{1}{2\sigma}(R_1+R_3)\,,\qquad q=\frac{1}{2\sigma}(R_1-R_3)\,,
\ee
where we have defined
\be
R_1\equiv\sqrt{\rho^2+(z+\sigma)^2}\,,\qquad
R_3\equiv\sqrt{\rho^2+(z-\sigma)^2}\,.
\ee
(This choice of numbering would become clearer below.) With this 
$\tilde f$ and $w$, we can immediately deduce $f$, $\phi$ and $B$ 
using (\ref{f}), (\ref{phi}) and the second equation of (\ref{ft}), 
respectively.

Now, to ensure that the line element (\ref{metric}) has a chance of being 
asymptotically flat, the harmonic function $h$ is taken to be 
\cite{Emparan_Reall}
\be
\label{h}
h=\ln\left[R_1+(z+\sigma)\right].
\ee
After integrating (\ref{mu_rho}) and (\ref{mu_z}) to obtain $\mu$, 
we finally arrive at the line element:
\ba
{\rm d}s^2&=&-H^\frac{8}{4+3\alpha^2}{\rm d}t^2+H^\frac{8}{4+3\alpha^2}
\left[\frac{(\sigma p+m)^2-a^2q^2}{\sigma^2(p^2-q^2)}\right]^\frac{12}
{4+3\alpha^2}
\frac{1}{2K_0R_1}\left({\rm d}\rho^2+{\rm d}z^2\right)\cr
&&+H^{-\frac{4}{4+3\alpha^2}}\left\{\left[R_1+(z+\sigma)\right]^{-1}\rho^2
{\rm  d}\varphi^2+\left[R_1+(z+\sigma)\right]{\rm d}\psi^2\right\}\,,
\ea
where $K_0$ is a dimensionless constant to be determined below, and
\be
\label{H}
H=\frac{\sigma^2p^2-a^2q^2-m^2}{(\sigma p+m)^2-a^2q^2}\,.
\ee
The dilaton and two-form gauge field are respectively given by
\be
\label{phiB}
\phi=\frac{6\alpha}{4+3\alpha^2}\ln H\,,\qquad
B_{\varphi\psi}=\sqrt{\frac{12}{4+3\alpha^2}}\,\frac{2am
(\sigma p+m)(1-q^2)}{\sigma^2p^2-a^2q^2-m^2}\,.
\ee
The rod structure of this solution along the $z$-axis can be deduced following 
\cite{Emparan_Reall}, and is shown in Fig.~1(a). As can be seen, the choice 
of $h$ in (\ref{h}) has produced semi-infinite rods corresponding to the 
$\varphi$ and $\psi$ coordinates, a property of asymptotically flat 
space-times such as the five-dimensional Schwarzschild solution 
\cite{Emparan_Reall}. Furthermore, the rod structure corresponding to the 
time coordinate shows that there are two `black' objects with horizons at 
$z=\pm\sigma$. The fact that these rods have shrunk down to points indicates
that they are extremally charged. The one at $z=-\sigma$ has a horizon with 
topology $S^3$; thus it is an extremal black hole. On the other hand, the 
one at $z=\sigma$ has a horizon with topology $S^2\times S^1$, and so it 
is an extremal black ring.

\begin{figure}[t]
\begin{center}
\includegraphics{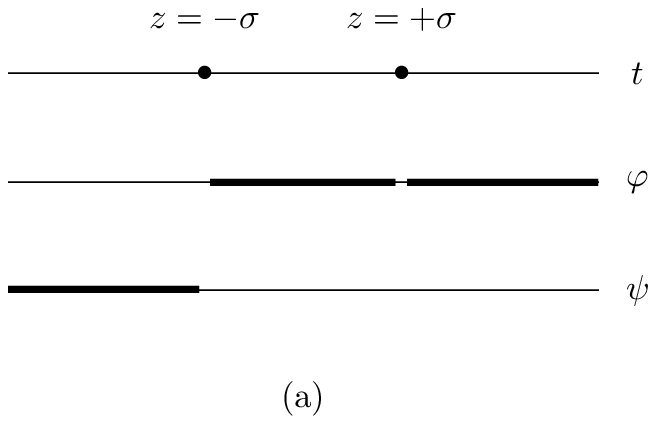}
\hskip1.5cm
\includegraphics{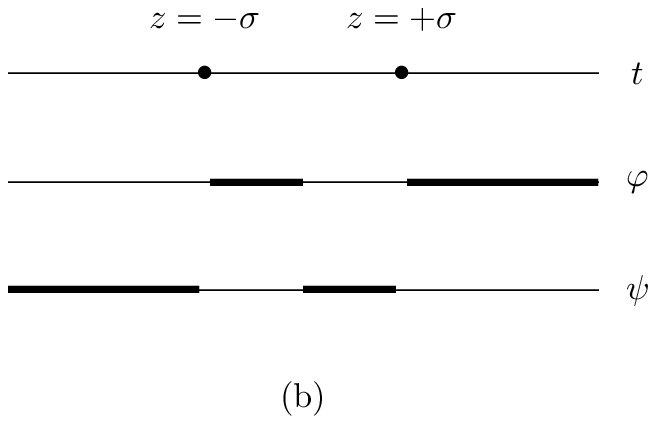}
\caption{Rod structures of (a) the extremal black hole / black ring system, 
and (b) the two-extremal black hole system.}
\end{center}
\end{figure}

To obtain instead a system describing two extremal black holes, we shall 
modify this solution to one with the rod structure as shown in Fig.~1(b). 
The difference is that part of the rod from $z=0$ to $\sigma$ has been
moved from the $\varphi$ to the $\psi$ coordinate. This is achieved quite 
simply by adding an appropriate harmonic function to $h$ in (\ref{h}). 
For the choice
\be
h=\ln\left(\frac{[R_1+(z+\sigma)](R_2-z)}{R_3-(z-\sigma)}\right),
\ee
where $R_2\equiv\sqrt{\rho^2+z^2}$, we arrive at the new solution:
\ba
\label{solution}
{\rm d}s^2&=&-H^\frac{8}{4+3\alpha^2}{\rm d}t^2+H^\frac{8}{4+3\alpha^2}
\left[\frac{(\sigma p+m)^2-a^2q^2}{\sigma^2(p^2-q^2)}\right]^\frac{12}
{4+3\alpha^2}\frac{Y_{12}Y_{23}}{R_1R_2R_3Y_{13}}\cr
&&\times\frac{1}{4K_0}\left({\rm d}\rho^2+{\rm d}z^2\right)
+H^{-\frac{4}{4+3\alpha^2}}\Bigg\{\frac{R_3-(z-\sigma)}
{[R_1+(z+\sigma)](R_2-z)}\,\rho^2{\rm d}\varphi^2\cr
&&+\frac{[R_1+(z+\sigma)](R_2-z)}{R_3-(z-\sigma)}\,{\rm d}\psi^2\Bigg\}\,,
\ea
where
\ba
Y_{12}&\equiv&\rho^2+(z+\sigma)z+R_1R_2\,,\cr
Y_{23}&\equiv&\rho^2+(z-\sigma)z+R_2R_3\,,\cr
Y_{13}&\equiv&\rho^2+(z^2-\sigma^2)+R_1R_3\,,
\ea
and with $H$, $\phi$ and $B_{\varphi\psi}$ unchanged as in (\ref{H}) and 
(\ref{phiB}). 

Now, the solution as written in the Weyl form (\ref{solution}) has the 
disadvantage that the $R_i$ contain square roots, making calculations 
somewhat cumbersome. Before proceeding any further, it would be useful 
to find alternative coordinates that would simplify the form of the
solution by getting rid of these square roots. 

If the rod structure had just two `characteristic points' where the rods
terminate, as in Fig.~1(a), then such 
a coordinate system will be provided by the prolate spheroidal 
coordinates (\ref{prolate}). But since the rod structure of our
final solution has three characteristic points (at $z_1=-\sigma$, $z_2=0$, 
and $z_3=\sigma$), a different coordinate system has to be adopted.
It turns out to be possible to introduce coordinates similar to that 
used in the standard form of the C-metric, whose rod structure also
has three characteristic points (see, e.g., \cite{Bonnor:1983}). The 
transformation in this case is~\cite{Hong}
\be
\rho=\frac{2\sigma\sqrt{xy(1-x^2)(y^2-1)}}{(x-y)^2}\,,\qquad
z=\frac{\sigma(x+y)(1-xy)}{(x-y)^2}\,,
\ee
where the new coordinates $(x,y)$ assume the finite range $0\leq x\leq 1$
and $-1\leq y\leq 0$. Under this transformation, $R_i$ take the simple 
algebraic form:
\be
R_1=-\sigma\frac{xy-x-y-1}{x-y}\,,\qquad
R_2=\sigma\frac{xy+1}{x-y}\,,\qquad
R_3=-\sigma\frac{xy+x+y-1}{x-y}\,.
\ee
This is the crucial simplification made possible by using the C-metric-type 
coordinates. 

In these new coordinates, the line element (\ref{solution}) becomes
\ba
\label{solution1}
{\rm d}s^2&=&-H^\frac{8}{4+3\alpha^2}{\rm d}t^2+H^\frac{8}{4+3\alpha^2}
\left[\frac{(\sigma p+m)^2-a^2q^2}{\sigma^2(p^2-q^2)}\right]^\frac{12}
{4+3\alpha^2}\cr
&&\times\frac{\sigma}{2K_0}\frac{1}{(x-y)^2}
\left(\frac{{\rm d}x^2}{x(1-x^2)}+\frac{{\rm d}y^2}{y(y^2-1)}\right)\cr
&&+H^{-\frac{4}{4+3\alpha^2}}\frac{2\sigma}{(x-y)^2}\{y(y^2-1)\,{\rm  d}\varphi^2
+x(1-x^2)\,{\rm d}\psi^2\}\,,
\ea
with $(p,q)$ given in terms of $(x,y)$ by 
\be
\label{pq}
p=\frac{1-xy}{x-y}\,,\qquad q=\frac{x+y}{x-y}\,.
\ee
The black holes at $(\rho,z)=(0,-\sigma)$ and $(0,\sigma)$ are respectively
located at $(x,y)=(0,-1)$ and $(1,0)$ in the new coordinate system, while
the origin $(\rho,z)=(0,0)$ is at $(1,-1)$. Asymptotic infinity is at $(0,0)$.

\newsection{Physical properties}

We now examine some physical properties of the solution (\ref{solution1}), 
beginning with its asymptotic structure. If we introduce coordinates
$(r,\theta)$ given by
\be
\label{changeofvariable}
x=\frac{2\sigma}{K_0r^2}\cos^2\theta\,,\qquad
y=-\frac{2\sigma}{K_0r^2}\sin^2\theta\,,
\ee
and take the limit $r\rightarrow\infty$, the line element becomes
\be
{\rm d}s^2=-{\rm d}t^2+{\rm d}r^2+r^2({\rm d}\theta^2+K_0\sin^2\theta\,{\rm d}
\varphi^2+K_0\cos^2\theta\,{\rm d}\psi^2)\,.
\ee
This is Minkowski space in spherical polar coordinates, with manifest
$U(1)\times U(1)$ rotational symmetry. However, to ensure 
that both $\varphi$ and $\psi$ have the usual periodicity of $2\pi$, we must 
set $K_0=1$. With this choice, (\ref{solution1}) becomes an asymptotically 
flat solution. From the asymptotic behavior of $g_{tt}$, we deduce that 
the mass of this solution is $\frac{16m}{4+3\alpha^2}$. 

When $m=0$, (\ref{solution1}) reduces to the vacuum space-time
\be
\label{C-metric}
{\rm d}s^2=-{\rm d}t^2+\frac{\sigma}{(x-y)^2}
\left(\frac{{\rm d}x^2}{2x(1-x^2)}+\frac{{\rm d}y^2}{2y(y^2-1)}
+2y(y^2-1)\,{\rm d}\varphi^2+2x(1-x^2)\,{\rm d}\psi^2\right).
\ee
The spatial part of this space-time is but the Euclideanized C-metric, 
for some specific choice of the mass and acceleration parameters.
The rod structure of (\ref{C-metric}) is given by that in Fig.~1(b), 
but without the two point sources belonging to the time coordinate. 
{}From the remaining rods corresponding to the $\varphi$ and $\psi$
coordinates, we see that there are two `outer' semi-infinite axes, and 
two `inner' finite ones. Since we have demanded that the outer axes have 
the usual periodicity, conical singularities necessarily appear along
the two inner ones \cite{Emparan_Reall}. It can be checked that they both 
have periods $4\pi$ \cite{Tan}, corresponding to struts in the space-time. 
As explained in \cite{Tan}, these struts are actually membranes, and have
the topology of a sphere. Furthermore, these two topological spheres are 
orthogonal to each other, since they are aligned along the two different 
rotational axes.

The three points, $z_1$, $z_2$, and $z_3$, where the two axes meet up are
fixed points of both $U(1)$'s. It is thus possible to introduce at these 
points, objects whose constant-radius surfaces have $S^3$ topology, such as 
black holes. Indeed, this background was first used in \cite{Emparan_Reall}
to construct an asymptotically flat three-black hole solution. It was
subsequently utilized in \cite{Tan} to construct a
two-black hole solution with equal charge-to-mass ratio (which was also
generalized to an $N$-black hole system). Here, we will show that
the solution (\ref{solution1}) actually corresponds to a two-black hole 
solution with opposite charge, i.e., a dihole.

Clearly, the solution is magnetically charged with respect to $H_{abc}$.
The asymptotic behavior of $B_{\varphi\psi}$ reveals that it
describes a dipole configuration, with a moment proportional to $ma$.
(Since changing the sign of $a$ is equivalent to reversing the orientation 
of the dipole, we may take $a\geq0$ without any loss of generality.)
The non-dilatonic case $\alpha=0$ is therefore a $U(1)\times U(1)$-symmetric 
five-dimensional generalization of Bonnor's magnetic dipole solution 
\cite{Bonnor}. 

To show that there are two black holes located at $(x,y)=(0,-1)$ 
and $(1,0)$, we have to change to suitable coordinates which blow up 
each of these regions, while sending the other off to infinity. Let us
concentrate on the left black hole at $(x,y)=(0,-1)$ first. An 
appropriate choice of coordinates is $(r,\theta)$ given by
\be
\label{changeofvariable1}
x=\frac{r^2}{2\sigma}\cos^2\theta\,,\qquad
y=-1+\frac{r^2}{\sigma}\sin^2\theta\,,
\ee
in the limit $a\rightarrow\infty$ such that $r\sin\theta$ 
and $r\cos\theta$ remain finite. We obtain:
\ba
\label{dilatonic}
{\rm
d}s^2&=&-\left(1+\frac{m}{r^2}\right)^{-\frac{8}{4+3\alpha^2}}{\rm d}t^2
+\left(1+\frac{m}{r^2}\right)^\frac{4}{4+3\alpha^2}\bigg[{\rm d}r^2\cr 
&&\hskip1.55in+r^2({\rm d}\theta^2+4\sin^2\theta\,{\rm
d}\varphi^2+\cos^2\theta\,{\rm d}\psi^2)\bigg]\,,\cr
\phi&=&-\frac{6\alpha}{4+3\alpha^2}\ln\left(1+\frac{m}{r^2}\right),\cr
B_{\varphi\psi}&=&\sqrt{{12\over{4+3\alpha^2}}}\,{2m}\cos^2\theta\,.
\ea
This is just the solution for an extremal dilatonic black hole 
\cite{Horowitz}, but with a conical singularity attached
to it along the $\varphi$-axis, with a period of $4\pi$. 
This conical singularity is an artifact of the background space-time
(\ref{C-metric}), since the latter has a conical singularity with
exactly the same period. Its presence will modify the calculation
of the ADM mass of this black hole, in that an extra factor of two will 
appear in the integral of energy-density over a spatial hypersurface 
\cite{Aryal}. The mass is obtained to be $\frac{8m}{4+3\alpha^2}$.

To recover the right black hole at $(x,y)=(1,0)$, the corresponding
transformation is 
\be
\label{changeofvariable2}
x=1-\frac{r^2}{\sigma}\cos^2\theta\,,\qquad
y=-\frac{r^2}{2\sigma}\sin^2\theta\,.
\ee
We obtain the solution:
\ba
{\rm
d}s^2&=&-\left(1+\frac{m}{r^2}\right)^{-\frac{8}{4+3\alpha^2}}{\rm d}t^2
+\left(1+\frac{m}{r^2}\right)^\frac{4}{4+3\alpha^2}\bigg[{\rm d}r^2\cr 
&&\hskip1.55in+r^2({\rm d}\theta^2+\sin^2\theta\,{\rm
d}\varphi^2+4\cos^2\theta\,{\rm d}\psi^2)\bigg]\,,\cr
\phi&=&-\frac{6\alpha}{4+3\alpha^2}\ln\left(1+\frac{m}{r^2}\right),\cr
B_{\varphi\psi}&=&\sqrt{\frac{12}{4+3\alpha^2}}\,{2m}\sin^2\theta\,.
\ea
There is now a conical singularity attached to the black hole
along the $\psi$-axis with a period of $4\pi$, as expected. 
Note that this black hole has opposite charge to the other one.
But its mass is also $\frac{8m}{4+3\alpha^2}$, giving a 
total mass of $\frac{16m}{4+3\alpha^2}$ for the system, in agreement with 
the result derived above.
 
In a similar vein, we may perform the transformation
\be
x=1-\frac{4r^2}{\sigma}\cos^2\theta\,,\qquad
y=-1+\frac{4r^2}{\sigma}\sin^2\theta\,,
\ee
to blow up the region around the origin $(x,y)=(1,-1)$. We obtain the 
line element
\be
{\rm d}s^2=-{\rm d}t^2+{\rm d}r^2+r^2({\rm
d}\theta^2+4\sin^2\theta\,{\rm
d}\varphi^2+4\cos^2\theta\,{\rm d}\psi^2)\,,
\ee
which is flat Minkowski space but with double the usual periodicity for 
both the angles $\varphi$ and $\psi$, as expected. There is nothing
special in this region, apart from the two conical singularities 
meeting at the $U(1)\times U(1)$ fixed point $r=0$.

There is another limit one could consider, namely the near-horizon
region of each black hole. For the left one, the transformation is 
the same as in (\ref{changeofvariable1}), except that instead of taking 
the limit of large $a$, we take the limit of small $r$. We obtain the 
line element:
\ba
\label{near-horizon}
{\rm d}s^2&=&g(\theta)^\frac{8}{4+3\alpha^2}
\Bigg[-\bigg(\frac{r^2}{Q}\bigg)^\frac{8}{4+3\alpha^2}{\rm d}t^2
+\bigg(\frac{Q}{r^2}\bigg)^\frac{4}{4+3\alpha^2}
({\rm d}r^2+r^2{\rm d}\theta^2)\Bigg]\cr 
&&+g(\theta)^{-\frac{4}{4+3\alpha^2}}
\bigg(\frac{Q}{r^2}\bigg)^\frac{4}{4+3\alpha^2}
 r^2(4\sin^2\theta\,{\rm d}\varphi^2+\cos^2\theta\,{\rm d}\psi^2)\,,
\ea
where we have defined $Q\equiv m(m+\sigma)/\sigma$ and
\be
g(\theta)=\sin^2\theta+\frac{a^2}{\sigma^2}\cos^2\theta\,.
\ee
The latter is a deformation factor which determines how the metric
deviates from spherical symmetry. A similar result may be obtained for 
the right black hole. Thus this shows that, even at finite 
distance, there are two, albeit distorted, black holes present in the 
space-time. The distortion suffered by each black hole is due 
to the forces exerted by the other black hole.

The two black holes are joined up by the inner $\varphi$- and $\psi$-axes.
The former is parameterized by $y=-1$ with $0<x<1$, while the latter
is parameterized by $x=1$ with $-1<y<0$. Note that each segment now has the 
topology of a disk \cite{Tan}. The two topological disks span orthogonal 
directions, and join up at the origin $(x,y)=(1,-1)$. 
By calculating the proper distance along these two segments, it may be 
seen that $a$ determines the separation of the black holes. For large 
$a$, this distance is proportional to $\sqrt{a}$. 
In the coincidence limit $a\rightarrow0$, the magnetic charge cancels out
and we obtain a five-dimensional dilatonic generalization of the 
Darmois solution \cite{Kramer}. The latter may be thought of as 
a superposition of two Schwarzschild black holes at the same point, but 
its physical interpretation remains rather obscure.

The non-dilatonic case $\alpha=0$ has to be treated separately, since the
proper distance calculated is always infinite. This is due to the well-known
fact that the Reissner--Nordstr\"om black hole has an infinite throat in
the extremal limit. However, $a$ still gives an indication of the separation
of the black holes, as can be seen, for example, from the dependence of 
the dipole moment on it.

Now recall that $K_0$ was chosen above so that the outer $\varphi$- and
$\psi$-axes have the usual periodicity. This implies that conical 
singularities necessarily appear along the inner $\varphi$- and $\psi$-axes. 
Indeed, the conical excess along both axes can be calculated to be
\be
\delta=2\pi\left[2\bigg(1+\frac{m^2}{a^2}\bigg)^\frac{6}{4+3\alpha^2}-1
\right],
\ee
which shows that there are two orthogonal disk-like struts joining the 
black holes. They provide the stress required to counterbalance the 
attraction between the black holes along the two orthogonal directions, 
thus yielding a static configuration. On the other hand, if we had chosen
\be
K_0=\frac{1}{4}\bigg(\frac{a^2}{m^2+a^2}\bigg)^\frac{12}{4+3\alpha^2},
\ee
the inner axes would have the usual periodicity, but conical singularities
would appear along the outer axes. It can be checked that both axes would
have a deficit angle of
\be
\delta=2\pi\left[1-\frac{1}{2}\bigg(\frac{a^2}{m^2+a^2}\bigg)^\frac{6}
{4+3\alpha^2}\right].
\ee
This may be interpreted as a pair of semi-infinite `cosmic membranes' 
(the higher-dimensional generalization of cosmic strings) pulling on the 
black holes in orthogonal directions to maintain equilibrium.

Apart from struts or cosmic membranes, it is possible to use a background 
magnetic field, appropriately aligned along the dipole, to provide the 
necessary forces to keep the system in equilibrium. Such a magnetic field 
can be introduced into the 
solution (\ref{solution1}) by means of a Harrison-type transformation. 
In four dimensions, such a transformation was generalized to dilaton 
gravity in \cite{Dowker:1993}. The corresponding transformation in the 
present case is\footnote{An analogous transformation for a magnetically
charged one-form gauge field in five dimensions can be found in 
\cite{Emparan:2001}.}
\ba
\label{Harrison}
\tilde f^{\prime}&=&\Lambda\tilde f\,,\cr
B_{\varphi\psi}^\prime&=&-\sqrt{\frac{12}{4+3\alpha^2}}\frac{1}{{\cal B}\Lambda}
\Bigg(1+\sqrt{\frac{4+3\alpha^2}{12}}{\cal B}B_{\varphi\psi}\Bigg)\,,\cr
{\rm e}^{\mu'}&=&\Lambda^\frac{8}{4+3\alpha^2}{\rm e}^\mu,
\ea
where
\be
\Lambda\equiv\Bigg(1+\sqrt{\frac{4+3\alpha^2}{12}}{\cal B}
B_{\varphi\psi}\Bigg)^2+{\cal B}^2\rho^2\tilde f^{-2},
\ee
and ${\cal B}$ is a constant that determines the background magnetic 
field strength. 
The proof of (\ref{Harrison}) involves writing this transformation
in terms of $(\tilde f^{\prime},w^\prime)$, and showing that it is
a solution to (\ref{fteqn}) and (\ref{weqn}). The behavior of $\mu$
under this transformation can then be deduced from (\ref{mu_rho})
and (\ref{mu_z}).

Applying the transformation (\ref{Harrison}) to the dihole solution, 
we obtain a solution that again takes the form (\ref{solution1}), 
(\ref{phiB}), but with
\ba
\label{newsolution}
H&=&\bigg\{(\sigma^2p^2-a^2q^2-m^2)+4{\cal B}am(\sigma p+m)(1-q^2)\cr
&&\quad+{\cal B}^2(1-q^2)
\bigg[\Big((\sigma p+m)^2-a^2\Big)^2+a^2\sigma^2(p^2-1)(1-q^2)\bigg]\bigg\}
\Big/\Big((\sigma p+m)^2-a^2q^2\Big)\,,\cr
B_{\varphi\psi}&=&\sqrt{\frac{12}{4+3\alpha^2}}\,\bigg\{2am(\sigma p+m)
+{\cal B}
\bigg[\Big((\sigma p+m)^2-a^2\Big)^2+a^2\sigma^2
(p^2-1)(1-q^2)\bigg]\bigg\}\,(1-q^2)\cr
&&\Big/\Big[H\Big((\sigma p+m)^2-a^2q^2\Big)\Big]\,.
\ea
An appropriate constant has been added to $B_{\varphi\psi}$ so that it 
reduces to the one in (\ref{phiB}) when ${\cal B}=0$. This solution 
describes a dihole configuration suspended in a background magnetic 
field. To see this, note that it has the same asymptotic limit (after 
performing the transformation (\ref{changeofvariable}) with $K_0=1$) 
as the solution
\be
\label{Melvin}
{\rm d}s^2=H^\frac{8}{4+3\alpha^2}(-{\rm
d}t^2+{\rm d}r^2+r^2{\rm d}\theta^2)+H^{-\frac{4}{4+3\alpha^2}}r^2
(\sin^2\theta\,{\rm d}\varphi^2+\cos^2\theta\,{\rm d}\psi^2)\,,
\ee
with
\ba
\label{Melvin2}
H&=&1+\frac{1}{4}{\cal B}^2r^4\sin^22\theta\,,\cr
\phi&=&\frac{6\alpha}{4+3\alpha^2}\ln H\,,\cr
B_{\varphi\psi}&=&\frac{1}{2}\sqrt{\frac{3}{4+3\alpha^2}}\,
H^{-1}{\cal B}r^4\sin^22\theta\,.
\ea
This is a five-dimensional analog of the dilatonic Melvin universe
\cite{Dowker:1993}. It describes a $U(1)\times U(1)$ symmetric magnetic 
field in an otherwise empty universe.

It is possible to find the near-horizon limit of this dihole solution using
the same transformation as above. For the left black hole, we again obtain 
a line element of the form (\ref{near-horizon}), but with the deformation 
factor:
\be
\label{distortion}
g(\theta)=\sin^2\theta+\frac{1}{\sigma^2}
\big[a+2{\cal B}m(m+\sigma)\big]^2\cos^2\theta\,.
\ee
Thus the presence of the background field will modify the shape of the
horizons, as to be expected. It can also be checked that the conical 
excess along the inner $\varphi$- and $\psi$-axes is now 
\be
\delta=2\pi\left[2\bigg(1+\frac{m^2}{a^2}\bigg)^\frac{6}{4+3\alpha^2}
\bigg(1+\frac{2{\cal B}m(m+\sigma)}{a}\bigg)^{-
\frac{12}{4+3\alpha^2}}-1\right].
\ee
By adjusting ${\cal B}$ appropriately, it is possible to set this to 
the natural value $2\pi$, i.e., the conical excess that is present in
(\ref{C-metric}) even when there are no black holes or background fields. 
This happens at the critical field strength
\be
{\cal B}_{\rm crit}=\frac{\sigma-a}{2m(m+\sigma)}\,,
\ee
and represents the situation in which the background magnetic field strength 
precisely cancels out the gravitational and magnetic attraction between the 
two black holes. Note that in this case, $g(\theta)=1$ and so the black 
holes are perfectly spherical, another consequence of the forces being 
balanced out. 

We remark that it is also possible to tune the background field strength to 
make $\delta=0$ and hence remove all the conical singularities from the 
space-time. However, this still leads to a non-trivial distortion 
factor in (\ref{distortion}), and can not be regarded as a natural 
situation in which the forces are canceling out.

Finally, we recall that in five dimensions, a solution that is 
magnetically charged with respect to the three-form field strength 
$H_{abc}$ may be dualized into another solution that is electrically 
charged with respect to a two-form field strength $F_{ab}$. 
This duality transformation takes the form
\be
\phi'=-\phi\,,\qquad F_{ab}={\rm e}^{\alpha\phi}(\ast H)_{ab}\,,
\ee
with the new solution extremizing the action
\be
\int{\rm d}^5x\sqrt{-g}\left(R-\frac{1}{2}\partial_a\phi'\partial^a\phi'
-\frac{1}{4}{\rm e}^{\alpha\phi'}F_{ab}F^{ab}\right).
\ee
Applying this transformation to (\ref{newsolution}), we obtain 
an electrically charged dihole solution in a background electric field.
The resulting electric potential is given by
\be
A_t=\sqrt{\frac{12}{4+3\alpha^2}}\Bigg\{\frac{2maq}{(\sigma p+m)^2-a^2q^2}
\Big[1-{\cal B}a(1-q^2)\Big]^2
-2{\cal B}q\Big[\sigma p-2m+{\cal B}ma(3-q^2)\Big]\Bigg\}\,.
\ee
In the asymptotic limit, it reduces to the electric dual of the
Melvin universe (\ref{Melvin}), (\ref{Melvin2}).

\newsection{Diholes with unbalanced charges}

Instead of starting with the four-dimensional Kerr solution, as in 
(\ref{4DKerr}), one can start with the Demia\'nski--Newman solution 
\cite{Demianski} which contains an additional NUT parameter $l$.
Repeating the procedure described in Sec.~3 and 4, we obtain the 
three-parameter solution (with $K_0=1$):
\ba
\label{NUT}
{\rm d}s^2&=&-H^\frac{8}{4+3\alpha^2}{\rm d}t^2+H^\frac{8}{4+3\alpha^2}
\left[\frac{(\sigma p+m)^2-(aq+l)^2}{\sigma^2(p^2-q^2)}
\right]^\frac{12}{4+3\alpha^2}\cr
&&\times\frac{\sigma}{2(x-y)^2}
\left(\frac{{\rm d}x^2}{x(1-x^2)}+\frac{{\rm d}y^2}{y(y^2-1)}\right)\cr
&&+H^{-\frac{4}{4+3\alpha^2}}\frac{2\sigma}{(x-y)^2}\{y(y^2-1)\,{\rm  d}\varphi^2
+x(1-x^2)\,{\rm d}\psi^2\}\,,
\ea
where now $\sigma\equiv\sqrt{m^2+a^2-l^2}$, and
\ba
\label{NUT2}
H&=&\frac{\sigma^2p^2-a^2q^2-m^2+l^2}
{(\sigma p+m)^2-(aq+l)^2}\,,\cr
\phi&=&\frac{6\alpha}{4+3\alpha^2}\ln H\,,\cr
B_{\varphi\psi}&=&\sqrt{\frac{12}{4+3\alpha^2}}\,\frac{2a
(m\sigma p+m^2-l^2)(1-q^2)+2l\sigma^2(p^2-1)q}
{\sigma^2p^2-a^2q^2-m^2+l^2}\,.
\ea
As usual, $(p,q)$ are related to $(x,y)$ by (\ref{pq}). This solution
clearly reduces to (\ref{solution1}), (\ref{H}) and (\ref{phiB}) 
when $l=0$. However, unlike the previous solution, it carries a net 
magnetic charge $-\sqrt{12\over4+3\alpha^2}4l$. We may assume $l\geq0$ 
without loss of generality, corresponding to a negatively charged solution. 

To recover the individual black holes, we change variables as in 
(\ref{changeofvariable1}) and (\ref{changeofvariable2}), to obtain
\ba
{\rm
d}s^2&=&-\bigg(1+\frac{m+l}{r^2}\bigg)^{-\frac{8}{4+3\alpha^2}}{\rm d}t^2
+\bigg(1+\frac{m+l}{r^2}\bigg)^\frac{4}{4+3\alpha^2}\bigg[{\rm
d}r^2\cr &&\hskip1.8in+r^2({\rm d}\theta^2+4\sin^2\theta\,{\rm
d}\varphi^2+\cos^2\theta\,{\rm d}\psi^2)\bigg]\,,\cr
\phi&=&-\frac{6\alpha}{4+3\alpha^2}\ln\bigg(1+\frac{m+l}{r^2}\bigg)\,,\cr
B_{\varphi\psi}&=&\sqrt{\frac{12}{4+3\alpha^2}}\,2(m+l)\cos^2\theta\,;
\ea
and
\ba
{\rm
d}s^2&=&-\bigg(1+\frac{m-l}{r^2}\bigg)^{-\frac{8}{4+3\alpha^2}}{\rm d}t^2
+\bigg(1+\frac{m-l}{r^2}\bigg)^\frac{4}{4+3\alpha^2}\bigg[{\rm
d}r^2\cr &&\hskip1.8in+r^2({\rm d}\theta^2+\sin^2\theta\,{\rm
d}\varphi^2+4\cos^2\theta\,{\rm d}\psi^2)\bigg]\,,\cr
\phi&=&-\frac{6\alpha}{4+3\alpha^2}\ln\bigg(1+\frac{m-l}{r^2}\bigg)\,,\cr
B_{\varphi\psi}&=&\sqrt{\frac{12}{4+3\alpha^2}}\,2(m-l)\sin^2\theta\,,
\ea
respectively. This shows that there is a black hole with mass $m+l$ at 
$(x,y)=(0,-1)$, and another one with mass $m-l$ at $(1,0)$. Thus, $l$ 
controls the distribution of mass (and hence charge) between the two 
black holes, although the total mass (charge) is still conserved. In 
the case $l=m$, one of the black holes disappears, leaving 
just a single black hole with mass $\frac{16m}{4+3\alpha^2}$. 

As with the $l=0$ case, the two black holes are held apart by struts
stretching between them. The conical excess along the inner $\varphi$- and
$\psi$-axes is
\be
\delta=2\pi\left[2\bigg(1+\frac{m^2-l^2}{a^2}\bigg)^\frac{6}{4+3\alpha^2}-1
\right].
\ee
Note that $\delta$ takes the natural value $2\pi$ when $m=l$, as to be 
expected when there is only one black hole left in the space-time. 

We may suspend the solution (\ref{NUT}), (\ref{NUT2}) in a background 
magnetic field ${\cal B}$, by applying the Harrison transformation 
(\ref{Harrison}) to it. The resulting solution again takes the form 
(\ref{NUT}) and (\ref{NUT2}), but with
\ba
H&=&\Bigg\{(\sigma^2p^2-a^2q^2-m^2+l^2)+
4{\cal B}\Big[a(m\sigma p+m^2-l^2)(1-q^2)+l\sigma^2(p^2-1)q\Big]\cr
&&\quad+{\cal B}^2(1-q^2)
\Bigg[\Big((\sigma p+m)^2-a^2-l^2\Big)^2+\bigg(a-\frac{2lq}{1-q^2}\bigg)^2\sigma^2(p^2-1)(1-q^2)\Bigg]\Bigg\}\cr
&&\Big/\Big((\sigma p+m)^2-(aq+l)^2\Big)\,,\cr
B_{\varphi\psi}&=&\sqrt{\frac{12}{4+3\alpha^2}}\,\Bigg\{2\Big[a
(m\sigma p+m^2-l^2)(1-q^2)+l\sigma^2(p^2-1)q\Big]\cr
&&\quad+{\cal B}(1-q^2)
\Bigg[\Big((\sigma p+m)^2-a^2-l^2\Big)^2+\bigg(a-\frac{2lq}{1-q^2}\bigg)^2\sigma^2
(p^2-1)(1-q^2)\Bigg]\Bigg\}\cr
&&\Big/\Big[H\Big((\sigma p+m)^2-(aq+l)^2\Big)\Big]\,.
\ea
It has the same asymptotic limit as the Melvin universe (\ref{Melvin}),
(\ref{Melvin2}).

The conical excess along the inner $\varphi$- and $\psi$-axes is then
\be
\delta=2\pi\left[2\bigg(1+\frac{m^2-l^2}{a^2}\bigg)^\frac{6}{4+3\alpha^2}
\bigg(1+\frac{2{\cal B}(m\sigma+m^2-l^2)}{a}\bigg)
^{-\frac{12}{4+3\alpha^2}}-1\right],
\ee
which takes the natural value $2\pi$ when the background field assumes the
strength
\be
{\cal B}_{\rm crit}=\frac{\sigma-a}{2(m\sigma+m^2-l^2)}\,.
\ee
However, unlike the $l=0$ case above, it is not possible to let the outer 
$\varphi$- and $\psi$-axes continue taking on their natural periodicity 
of $2\pi$. Instead, they must take the values
\ba
\Delta\varphi&=&2\pi(1+{\cal B}l)^\frac{12}{4+3\alpha^2},\cr
\Delta\psi&=&2\pi(1-{\cal B}l)^\frac{12}{4+3\alpha^2},
\ea
which implies that there is a conical excess along the $\varphi$-axis,
and a conical deficit along the $\psi$-axis. This is due to the asymmetric 
distribution of the charge, and is similar to the situation in four 
dimensions \cite{Liang}.

\newsection{Discussion}

In this paper, we have developed a solution-generating technique and used
it to obtain a five-dimensional dihole solution of dilaton gravity coupled
to a two-form gauge field. The main properties of this solution were then
studied. In particular, it was shown that there are membrane-like conical 
singularities in the space-time keeping the black holes in static
equilibrium. We also presented a dihole solution suspended in a background
magnetic field, as well as one with unbalanced charges.

There are a few possible extensions of this work. For example, it would
be interesting to study its thermodynamic properties and find a microscopic 
description for it following the methods of \cite{Costa}, as was done in 
\cite{Emparan_Teo} in the four-dimensional 
case. However, this requires one to first obtain the non-extremal dihole 
solution, perhaps using a five-dimensional generalization of Sibgatullin's 
method \cite{Sibg}. Such a solution, even if it could be constructed, is 
likely to be very complicated like its four-dimensional counterpart 
\cite{Emparan_Teo}.

There is also the question how the five-dimensional dihole solution can be 
embedded and interpreted in string or M-theory. When (\ref{solution1}) is 
uplifted to ten dimensions in the usual way, it appears to describe 
a non-standard type of 5-brane -- anti-5-brane configuration. Normally, 
a 5-brane is assumed to have $SO(4)$ symmetry in the transverse directions, 
and a space-time containing two 5-branes would therefore be expected to 
have maximal $SO(3)$ symmetry. However, our solution has only 
$U(1)\times U(1)$ symmetry, and this is tied in to the presence of the 
two conical membranes stretching between the 5-brane and anti-5-brane.

This leads one to wonder if there exists another dihole solution with 
$SO(3)$ instead of $U(1)\times U(1)$ symmetry. However, since such a 
space-time will not belong to the generalized Weyl class \cite{Emparan_Reall},
the procedure used to obtain the dihole solutions in this paper would simply 
not apply. In view of this, a radically different approach might be needed
(see, e.g., \cite{Charmousis}).

It would also be of interest to try to find dihole solutions in six or 
higher dimensions. Unfortunately, such solutions will not belong to the 
generalized Weyl class, since black holes in $D\geq6$ dimensions do not
admit $D-2$ commuting Killing vectors \cite{Emparan_Reall}. Again, 
a different approach would be needed to construct such solutions, and 
we do not have anything to add in this respect.

Finally, recall that in the derivation of the dihole solution in Sec.~3, 
an intermediate solution consisting of an extremal black hole surrounded
by an extremal black ring was constructed. It is possible to check that
they carry opposite charges, which cancel out to give a system with
zero net charge. This extremal black ring is in fact equivalent, in
the non-dilatonic case, to one obtained recently in \cite{Maeda}, and 
it has a singular event horizon with vanishing area. It might be 
worthwhile to study this system in its own right.

\bigskip\bigskip

{\renewcommand{\Large}{\normalsize}
}
\end{document}